\documentstyle[aps,prc]{revtex}

\begin{document}
\draft
\title{A general numerical solution of dispersion relations for the nuclear optical
model}
\author{Roberto Capote\cite{capo}$^{,1,2}$, Alberto Molina\cite{alberto}$^{,1}$ and
Jos\'{e} Manuel Quesada\cite{jmanuel}$^{,1}$}
\address{$^1$Departamento de F\'{i}sica At\'{o}mica, Molecular y Nuclear,\\
Universidad de Sevilla, Facultad de F\'{i}sica, Apdo 1065, 41080 Sevilla,\\
Espa\~{n}a\\
$^2$Centro de Estudios Aplicados al Desarrollo Nuclear, Calle 30 \# 502, \\
Miramar, La Habana, Cuba}
\date{\today}
\maketitle

\begin{abstract}
A general numerical solution of the dispersion integral relation between the
real and the imaginary parts of the nuclear optical potential is presented.
Fast convergence is achieved by means of the Gauss-Legendre integration
method, which offers accuracy, easiness of implementation and generality for
dispersive optical model calculations. The use of this numerical integration
method in the optical-model parameter search codes allows for a fast and
accurate dispersive analysis.
\end{abstract}


For many years the evaluation of reaction cross section and elastic
scattering data has relied on the use of the optical model. A significant
contribution during the last two decades can be considered the work of
Mahaux and co-workers on dispersive optical-model analysis\cite{Mahaux DOM
1,Mahaux DOM 2,Mahaux DOM 3}. The unified description of nuclear mean field
in dispersive optical-model is accomplished by using a dispersion relation
(DR), which links the real and absorptive terms of the optical model
potential (OMP). The dispersive optical model provides a natural extension
of the optical model derived data into the bound state region. In this way a
physically self-consistent description of the energy dependence of the OMP
is obtained and the prediction of single-particle, bound state quantities
using the same potential at negative energies became possible. Moreover
additional constraint imposed by DR helps to reduce the ambiguities in
deriving phenomenological OMP parameters from the experimental data. The
dispersive term of the potential $\triangle V(E)$ can be written in a form
which is stable under numerical treatment\cite{Delaroche}, namely:

\begin{equation}
\triangle V(E)=\frac 2\pi (E-E_F){\int_{E_F}^\infty }\frac{W(E^{\prime
})-W(E)}{(E^{\prime }-E_F)^2-(E-E_F)^2}dE^{\prime }  \label{integral}
\end{equation}
where $E$ is the incident projectile energy and $E_F$ is the Fermi energy
for the target system.

The imaginary OMP can be written as the sum of volume and surface
contributions: $W(E)=W_V(E)+W_S(E)$. It is useful to represent the variation
of surface $W_S(E)$ and volume absorption potential $W_V(E)$ depth with
energy in functional forms suitable for the DR optical model analysis. An
energy dependence for the imaginary volume term has been suggested in
studies of nuclear matter theory:

\begin{equation}
W_V(E)=\frac{A_V(E-E_F)^n}{(E-E_F)^n+(B_V)^n}  \label{volumen}
\end{equation}
where $A_V$ and $B_V$ are undetermined constants. Mahaux and Sartor\cite
{Mahaux DOM 3} have suggested $n=4$, while Brown and Rho\cite{Brown}, $n=2$.
An energy dependence for the imaginary-surface term has been suggested by
Delaroche {\it et al}\cite{Delaroche} to be:

\begin{equation}
W_S(E)=\frac{A_S(E-E_F)^m}{(E-E_F)^m+(B_S)^m}\exp (-C_S|E-E_F|)
\label{surface2}
\end{equation}
where $m=4$ and $A_S,B_S$ and $C_S$ are undetermined constants. Other energy
functionals are suggested in the literature \cite{Mahaux DOM 2,Weisel}.
According to equations (\ref{volumen}) and (\ref{surface2}) the imaginary
part of the OMP is assumed to be zero at $E=E_F$ and nonzero everywhere
else. A more realistic parametrization of $W_V(E)$ and $W_S(E)$ forces these
terms to be zero in some region around the Fermi energy. A physically
reasonable energy for defining such a region is the average energy of the
single-particle states $E_P$\cite{Mahaux DOM 1}. We can define the
offset-energy $E_{of}$ for both the particle and the hole region by the
relation: 
\begin{equation}
E_{of}=E_P-E_F
\end{equation}

Therefore a new definition for imaginary part of the OMP\ can be written as:

\begin{eqnarray}
W_V(E) &=&0,\text{ for }E_F<E<E_p  \label{WV2} \\
W_V(E) &=&\frac{A_V(E-E_P)^n}{(E-E_P)^n+(B_V)^n},\text{ for }E_P<E
\label{WV3}
\end{eqnarray}
and likewise for surface absorption. The symmetry condition $W(2E_F-E)=W(E)$
is used to define imaginary part of the OMP for energies below the Fermi
energy. We assume below that equations (\ref{volumen}) and (\ref{surface2})
are modified according to (\ref{WV2}) and (\ref{WV3}).

In a recent work\cite{AnalytDOM}, analytical solutions of the dispersion
integral relation between the real and imaginary parts of the nuclear
optical model were proposed. The main reason for such effort was the slow
convergence of the Simpson integration method (SIM) for solving the DR
integral (\ref{integral}). The authors needed 6000 steps in total (50 keV
each) to calculate (\ref{integral}) up to 300 MeV. Beyond 300 MeV, the
integral was approximated by assuming that the numerator of its integrand
become constant-valued\cite{AnalytDOM}. Obviously this lengthen very much
the OMP calculations, so they derived approximate analytical solutions of
the integral (\ref{integral}) for usual forms of the imaginary OMP,
including those given by (\ref{volumen}) and (\ref{surface2}). However the
approximate analytical solution recently derived\cite{AnalytDOM} depends, of
course, on assumed functional form of the imaginary part of the OMP.

In this work we propose a fast and general numerical solution for
calculation of the DR integral (\ref{integral}), based on Gauss-Legendre
integration method (GLIM) \cite{HANDBOOK}. Integral in a finite interval $%
(a,b)$ can be converted by linear transformation to the integral in the
interval $(-1,1)$ and then approximatted according to the following formula:

\begin{equation}
{\int_{-1}^1}f(x)dx={\sum }_{i=1}^NW_i^Nf(X_i^N)  \label{GAUSS}
\end{equation}
where $W_i^N$ and $X_i^N$ are weights and abscissas corresponding to the $N$
point Gauss-Legendre integration method \cite{HANDBOOK}. In this work we
were using N=10 for each interval. At the first look of equation (\ref
{integral}) we could be tempted to use Gauss-Laguerre integration method\cite
{HANDBOOK}. However beyond some energy cut-off $E_{cut}$, such that $%
E_{cut}>>E$, the integral can be very well approximated by assuming that the
numerator of its integrand becomes constant-valued. In this case we obtain
the following relation:

\begin{equation}
\triangle V_{res}(E)=\frac 2\pi (E-E_F){\int_{E_{CUT}}^\infty }\frac{%
W(E^{\prime })-W(E)}{(E^{\prime }-E_F)^2-(E-E_F)^2}dE^{\prime }=\frac{W(E)}%
\pi \log 
{E_{cut}-(E-E_F) \overwithdelims() E_{cut}+(E-E_F)}%
\label{ResInt}
\end{equation}
Therefore we can apply GLIM from $E_F$ up to $E_{cut}$ for a finite
interval. This option was shown to be more accurate for a given problem.
This method is suitable for practically any type of DR integrand function.
Moreover the implementation of this method in optical model parameter search
code is straightforward. The method is compared with SIM and, for the
surface dispersive contribution, with the results of analytical integration%
\cite{AnalytDOM}. The parameters of the dispersive OMP used in our tests
correspond to the ''partially constrained set'' derived in Ref.\cite{Weisel
96} to describe neutron scattering on $^{209}$Bi up to 80 MeV incident
energy (see Table II,\cite{Weisel 96}). For $^{209}$Bi the Fermi energy is
equal to -5.98 MeV and the offset energy was taken as 3.26 MeV, assuming the
average energy of the particle states is the same of $^{208}$Pb\cite{Weisel
96}.

In order to attain a good accuracy in the numerical integration (\ref
{integral}), attention must be paid to the behavior of the integrand,

\begin{equation}
I_{E,E_F}(E^{\prime })=\frac{W(E^{\prime })-W(E)}{(E^{\prime
}-E_F)^2-(E-E_F)^2}  \label{integrand}
\end{equation}

The integrand $I_{E,E_F}(E^{\prime })$ for dispersive surface correction ($%
W_S(E)$ given by equation (\ref{surface2}) ) is shown in Figure 1. As a
result of the exponential damping in the surface OMP the integrand $%
I_{E,E_F}(E^{\prime })$ decreases sharply above 5-6 MeV. This behavior
suggests that separation of the integration interval is desirable in order
to improve the accuracy of the numerical integration method. It should be
stressed that exponential damping contained in functional form (\ref
{surface2}) reflects the physical fact that surface absorption is expected
to decrease fast as the incident energy is increased. This behavior is
observed in almost any phenomenological OMP analysis, so dividing
integration interval is not a recipe for functional form (\ref{surface2}),
but should be valid practically for any imaginary surface absorption energy
functional. We decided to divide the interval in two parts from $E_F$ up to $%
E_{int}=10MeV$ and then from $E_{int}$ up to some cut-off energy in the
integral. The $E_{int}$ position was chosen in order to guarantee monotonous
behavior of the integrand in the interval from $E_{int}$ up to infinity.
Practically it is enough to select $E_{int}$ slightly larger than the energy
of the maximum value of the integrand. The cut-off energy $E_{cut}$ in the
DR integral for dispersive surface correction was taken at 1000 MeV. Beyond
1000 MeV we used equation (\ref{ResInt}) to estimate contribution from the
tail of the integral. This residual value was added to both numerical
methods. The dispersive surface correction $\triangle V_S(E)$ calculated by
GLIM (using 20 integration points) and SIM (100 keV step, more than 10000
integration points) were compared with analytical solution from Ref.\cite
{AnalytDOM}. Maximum deviation between any of the three methods was less
than 1 KeV in the whole energy range from $E_F$ up to 150 MeV energy.

A similar behavior of the integrand $I_{E,E_F}(E^{\prime })$ is observed for
the case of the dispersive volume correction ($W_V(E)$ given by equation (%
\ref{volumen}) ). However the integrand $I_{E,E_F}(E^{\prime })$ is smoother
and has longer tail than was the case for surface correction. The reason is
that no exponential damping is introduced here. Similarly to what we did
before, we divided the interval in two parts from $E_F$ up to $E_{int}=40MeV$
and then from $E_{int}$ up to cut-off energy $E_{cut}$. Beyond 1000 MeV we
used equation (\ref{ResInt}) to estimate contribution from the tail of the
integral. The integration was performed by GLIM and SIM. An excellent
agreement is achieved between both numerical methods.

\bigskip

In conclusion, we have proposed an accurate and general numerical method for
calculating dispersive contribution to the real part of the optical model
potential. The method is very fast and thanks to its computational
simplicity should be easy to implement in current generation of the optical
model parameter search codes. We stress that, unlike analytical solution,
any energy functional representation of the imaginary part of the OMP can be
treated in our approach. These properties make the method very useful for
introduction of the dispersive optical model relation in large scale nuclear
data calculations.

\smallskip

\acknowledgements
The authors are very indebted to Dr.G.Weisel for discussion about different
integrations methods, comments and recommendations, as well as Prof.M.Lozano
for his comments. One of the authors (R.C.) acknowledges support from the
Ministerio de Educaci\'{o}n, Ciencia y Cultura de Espa\~{n}a.

\newpage

\begin{figure}[tbp]
\caption{Energy dependence of the DR integrand $I_{E,E_F}(E^{\prime })$ for
the dispersive surface contribution of the imaginary OMP\newline
(short dottted line) $E = E_F+0.25 MeV$, (dash dotted line) $E = 0$, (dotted
line) $E = 10 MeV$ \newline
(dashed line) $E = 30 MeV$, (solid line) $E = 50 MeV$}
\label{Figure.1}
\end{figure}

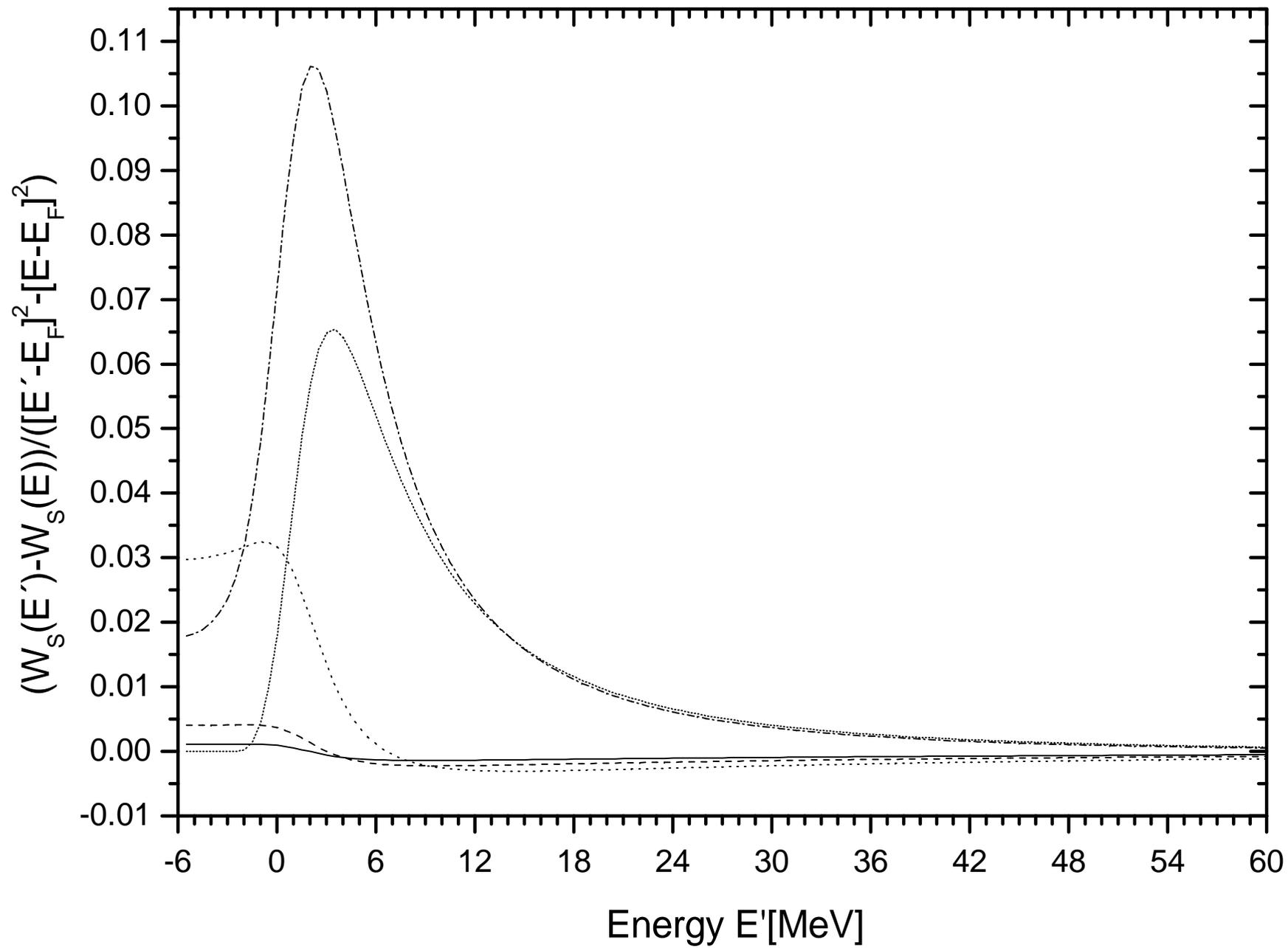
\end{document}